\documentclass[lettersize,journal]{IEEEtran}
\usepackage{cite}
\usepackage{siunitx} 
\usepackage{amsmath,amssymb,amsfonts}
\usepackage{graphicx}
\usepackage{listings}
\usepackage{textcomp}
\usepackage{xcolor}
\usepackage[most]{tcolorbox} 
\usepackage{algorithm}
\usepackage{algpseudocode}
\usepackage{tikz}
\usepackage{subcaption}
\usepackage{adjustbox}
\usepackage{caption}
\usepackage{setspace}
\usepackage{pgf-pie}  
\usepackage{makecell}
\usepackage{multirow}
\usepackage{url}
\usepackage{siunitx}
\usepackage{balance}
\usepackage{pgfplots}
\usepackage{tabularx}
\usepackage{pifont}
\usepackage{pgfplots}
\usepackage{tabularx}
\usepackage{graphicx}
\usepackage{booktabs}
\usepackage{enumitem}

\def\BibTeX{{\rm B\kern-.05em{\sc i\kern-.025em b}\kern-.08em
    T\kern-.1667em\lower.7ex\hbox{E}\kern-.125emX}}

\definecolor{codegreen}{rgb}{0,0.6,0}
\definecolor{codegray}{rgb}{0.5,0.5,0.5}
\definecolor{codepurple}{rgb}{0.58,0,0.82}
\definecolor{backcolour}{rgb}{0.95,0.95,0.92}
\definecolor{codered}{rgb}{0.91,0.58,0.53}

\definecolor{softblue}{RGB}{102,178,255}
\definecolor{softred}{RGB}{255,153,153}
\definecolor{softgreen}{RGB}{153,255,153}
\definecolor{softyellow}{RGB}{255,255,153}
\definecolor{softcyan}{RGB}{153,255,255}
\definecolor{softmagenta}{RGB}{255,153,255}
\definecolor{softorange}{RGB}{255,200,150}
\lstdefinestyle{mystyle}{
    backgroundcolor=\color{backcolour},
    commentstyle=\color{black},
    keywordstyle=\color{black},
    numberstyle=\scriptsize\color{codegray},
    stringstyle=\color{black},
    basicstyle=\ttfamily\scriptsize, 
    breakatwhitespace=false,
    breaklines=true,
    captionpos=b,
    keepspaces=true,
    numbers=left,
    numbersep=5pt,
    showspaces=false,
    showstringspaces=false,
    showtabs=false,
    tabsize=2
}

\lstdefinestyle{mystyle_1}{
    backgroundcolor=\color{codered},
    commentstyle=\color{codegreen},
    keywordstyle=\color{black},
    numberstyle=\scriptsize\color{codered},
    stringstyle=\color{codered},
    basicstyle=\ttfamily\scriptsize, 
    breakatwhitespace=false,
    breaklines=true,
    captionpos=b,
    keepspaces=true,
    numbers=left,
    numbersep=5pt,
    showspaces=false,
    showstringspaces=false,
    showtabs=false,
    tabsize=2
}
\newcommand{\tool}{\textsc{VistaFuzz}}

\newif\ifmarkrevision
\markrevisionfalse

\begin{document}

\title{Harnessing LLMs for Document-Guided Fuzzing of Python Libraries}

\author{
  Bin Duan$^{1}$,
  Tarek Mahmud$^{2}$,
  Meiru Che$^{3}$,
  Yan Yan$^{4}$,
  Naipeng Dong$^{1}$,
  Dan Dongseong Kim$^{1}$,
  Guowei Yang$^{1*}$ \\
  $^{1}$School of Electrical Engineering and Computer Science, The University of Queensland, Australia \\
  $^{2}$Department of Computer Science, Texas State University, USA \\
  $^{3}$College of Information and Communications Technology, Central Queensland University, Australia \\
  $^{4}$Department of Computer Science, University of Illinois Chicago, USA \\
  \{b.duan, n.dong, dan.kim, guowei.yang\}@uq.edu.au, \\
  {tarek\_mahmud@txstate.edu}, {m.che@cqu.edu.au}, {yyan55@uic.edu}\\
}


\maketitle

\begingroup
\renewcommand\thefootnote{*}
\renewcommand{\footnoterule}{}
\footnotetext{Corresponding author.}
\endgroup
\begin{abstract}
Python libraries provide the foundation for deep learning, scientific computing, data analysis, and computer vision, and their reliability is critical for downstream applications. This paper introduces \tool, a document-guided fuzzing technique for Python library APIs. \tool\ uses a locally served open-sourced large language model (LLM) to parse heterogeneous API documentation into standardized parameter specifications, validates them against runtime signatures, and normalizes recurring inter-parameter dependencies into executable relationships for valid test-input generation. We evaluate \tool\ on 7,718 APIs across twelve Python libraries. Inter-parameter relationships occur in 40.1\% of the tested APIs, and disabling relationship resolution reduces the valid generation rate on such APIs from above 95\% to 32.1\%--52.3\%. \tool\ reports 74 issues, of which 47 have been confirmed by developers as real bugs and 38 have already been fixed. On PyTorch and TensorFlow, we compare \tool\ with API-level fuzzers, an LLM-based fuzzer, and the same LLM prompted to generate test inputs directly, and report API applicability, effectiveness on the commonly supported APIs under the same per-API budget, and bug detection under the same oracles separately. \tool\ supports more APIs and covers more lines than the API-level fuzzers, is comparable with the LLM-based fuzzer on the commonly supported APIs, on which the same LLM prompted directly covers 23\% and 25\% fewer lines, and detects 12 of 27 known bugs, as many as the LLM-based fuzzer and more than the other baselines.
\end{abstract}

\begin{IEEEkeywords}
Fuzzing, Python Libraries, Large Language Models
\end{IEEEkeywords}

\section{Introduction}
\label{sec:intro}

Python libraries such as PyTorch~\cite{paszke2019pytorch}, TensorFlow~\cite{abadi2016tensorflow}, NumPy~\cite{harris2020array}, SciPy~\cite{virtanen2020scipy}, scikit-learn~\cite{pedregosa2011scikit}, and OpenCV~\cite{bradski2000opencv} provide fundamental functionality for deep learning, scientific computing, data analysis, and computer vision, and defects in them propagate silently into downstream applications~\cite{zhang2018empirical,islam2019comprehensive}. Their APIs accept heterogeneous inputs involving tensors, arrays, scalar values, shapes, optional arguments, and dependencies among parameters. Generating executable test cases for these APIs is challenging because each invocation must satisfy both general Python semantics and API-specific constraints on types, dimensions, value ranges, and parameter relationships~\cite{xie2022docter,wei2022free}.

Traditional library fuzzers construct valid API invocations using manually defined rules, seed programs, or mined code examples~\cite{xie2022docter,wei2022free}, or exercise libraries indirectly through generated neural networks~\cite{pham2019cradle,wang2020lemon,gu2022muffin,guo2020audee} and relational API equivalences~\cite{deng2022fuzzing,duan2025xamt,duan2026testing}. Recent LLM-based approaches offer an advanced direction: TitanFuzz~\cite{deng2023large} synthesizes and mutates programs for deep-learning libraries, FuzzGPT~\cite{deng2024large} uses historical bug-triggering programs to generate unusual API usages, and Fuzz4All~\cite{xia2024fuzz4all} extends LLM-based generation to diverse input languages. These approaches primarily generate complete test programs, leaving API constraints implicit in code, examples, or prompts. The knowledge sources of the earlier library fuzzers---mined corpora and learned parsing rules---are library-specific, and these fuzzers have been evaluated mainly on PyTorch and TensorFlow. More recent work moves beyond these two libraries or towards explicit constraints: FUTURE~\cite{li2025future} transfers historical bug-triggering code to newly introduced libraries with fine-tuned LLMs, and FD-FACTORY~\cite{zheng2026fdfactory}, published after our conference paper~\cite{duan2025harnessing}, lets an LLM analyze documentation into constraints that guide the generation of a fuzz driver for each API.
API documentation, by contrast, explicitly describes parameter types, shapes, value ranges, defaults, and dependencies for virtually every released library. DocTer~\cite{xie2022docter} demonstrates that such documentation can be translated into executable constraints. Doing so with one pipeline across libraries faces three challenges. \textit{(C1) Heterogeneous documentation:} formats and completeness differ from library to library, so parsing rules learned for one library are tied to its format. \textit{(C2) Inter-parameter dependencies:} they are described in free text and worded differently across libraries, and inputs that ignore them are often rejected as invalid. \textit{(C3) Library-specific invocation:} each library has its own calling conventions and input objects, and a specification extracted by an LLM can disagree with the actual signature. Existing document-guided approaches handle these per library or per API: DocTer relies on rules learned for the documentation formats of specific libraries, and FD-FACTORY realizes LLM-extracted constraints in a separate LLM-written driver for each API. A single specification that is filled from heterogeneous documentation, checked against the runtime signature, and executed by one generic generator across libraries remains insufficiently explored.

To address this gap, we present \tool, a documentation-driven fuzzing approach for Python library APIs. For each API, \tool\ collects its documentation and runtime signature, and uses an LLM to convert them into standardized parameter specifications (\textit{ParamSpec}s) that capture parameter types, shapes, value ranges, default values, individual constraints, and inter-parameter relationships normalized into three executable patterns (shape following, rank-bounded axes, and type following). It validates each specification against the runtime signature and materializes it into library-native arrays, tensors, scalars, and structured arguments for test execution. The LLM is invoked once per API using a locally served open-sourced model with greedy decoding, so extraction is deterministic, reproducible, and free of per-test inference cost. Thus, one schema and one prompt replace library-specific parsing rules (C1), a generic resolver executes the three relationship patterns in every library (C2), and signature validation together with a thin library-facing layer for documentation collection and materialization absorbs what is specific to a library (C3).
\tool\ employs two complementary test oracles: a crash oracle that detects abnormal process termination and unexpected exceptions on constraint-satisfying inputs, and a NaN oracle that flags non-finite outputs inconsistent with documented behavior on constraint-satisfying inputs.

We evaluate \tool\ on twelve Python libraries: PyTorch, TensorFlow, JAX, Keras, PaddlePaddle, OneFlow, MindSpore, Chainer, NumPy, SciPy, scikit-learn, and OpenCV. Inter-parameter relationships appear in 40.1\% of the tested APIs, and disabling relationship resolution reduces the valid generation rate on such APIs from above 95\% to 32.1\%--52.3\%. Across 7,718 APIs, \tool\ reports 74 issues, of which 47 have been confirmed by developers as real bugs and 38 have already been fixed. On PyTorch and TensorFlow, we compare \tool\ with API-level fuzzers (FreeFuzz, DocTer, and ACETest), an LLM-based fuzzer (TitanFuzz), and the same LLM prompted to generate test inputs directly from the same documentation and signatures; on MindSpore and OneFlow, we compare it with FUTURE. We report API applicability, effectiveness on the APIs supported by all tools under the same per-API budget, and bug detection under the same oracles separately. \tool\ supports more APIs and covers more lines than the three API-level fuzzers and is comparable with TitanFuzz on the commonly supported APIs, on which the same LLM prompted directly covers 23\% and 25\% fewer lines. On a benchmark of 27 known bugs, \tool\ and TitanFuzz each detect 12, and the other baselines 3--8. The same core extraction and generation pipeline is applied across all twelve libraries, with a median of 112 lines of library-specific code for each library added beyond OpenCV. We release the implementation of \tool~\cite{ours}.

This article extends our IEEE ICSME 2025 paper~\cite{duan2025harnessing}, which focused on document-guided fuzzing of OpenCV. The methodological extension is the transition from an OpenCV-oriented representation to a cross-library \textit{ParamSpec} that combines documentation with runtime signatures and represents recurring dependencies in executable forms. Replacing GPT-4 with a locally served open-sourced model is primarily a reproducibility improvement. Experimentally, the study expands from 330 APIs in one library to 7,718 APIs across twelve libraries: the eleven added libraries account for 7,388 of the tested APIs, 65 of the 74 reported issues, and 38 of the 47 confirmed bugs. The article also adds controlled comparisons with six baselines, a benchmark of 27 known bugs, component ablations, an accuracy analysis of the extracted specifications on 360 labeled APIs, and analyses of dependency prevalence, generalizability, and cost.

\begin{figure}[t!]
  \centering
  \includegraphics[width=0.48\textwidth]{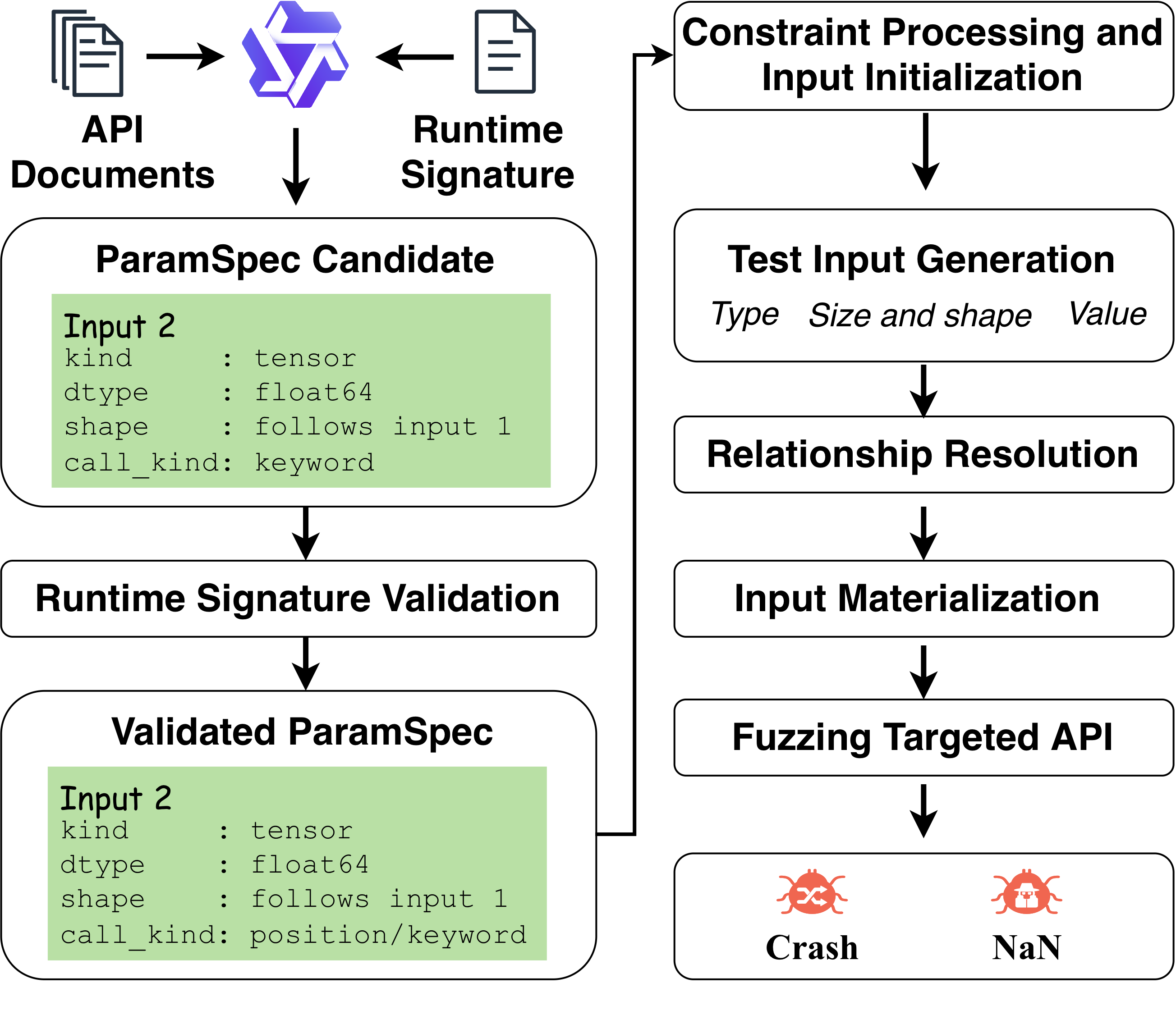}
  \caption{Overview of \tool}
  \label{fig:overview}
\end{figure}

\section{Approach}
\label{sec:approach}

Figure~\ref{fig:overview} presents the workflow of \tool. For each target API, \tool\ collects its official documentation and runtime signature and provides them to an LLM. The LLM converts heterogeneous natural-language descriptions into structured parameter specifications, including individual constraints and the inter-parameter relationships supported by \tool. The extracted specifications are validated against the runtime signature and then used to initialize and generate API arguments. During generation, \tool\ varies parameter types, sizes, and values, resolves the supported relationships using already generated arguments, and materializes the resulting values into the runtime objects required by the target library. Finally, \tool\ invokes the target API and applies the crash and NaN oracles.

\subsection{Standardizing API Information}
\label{sec:standardization}

Python libraries use heterogeneous documentation formats and parameter conventions. Some APIs provide complete signatures and detailed descriptions, whereas others omit types, shapes, valid ranges, defaults, or relationships among parameters. \tool\ therefore converts the documentation and runtime information of each API into a standardized representation before generating test inputs.

\subsubsection{Information Collection}

For each target API, \tool\ collects its official documentation and runtime signature. The documentation provides natural-language descriptions of parameter types, shapes, value domains, defaults, and relationships. The runtime signature provides concrete invocation information, including parameter names, positions, required and optional parameters, default values, and calling forms.

The two sources are complementary. Documentation contains richer semantic information but may be incomplete, inconsistent, or expressed using library-specific terminology. Runtime signatures reliably describe how an API is invoked but generally provide limited information about value domains, shapes, and relationships among parameters.

\subsubsection{Prompt Construction and LLM-Based Extraction}

\tool\ uses an LLM to convert the collected information into structured parameter specifications. For each target API, the prompt consists of four parts: (1) a task instruction defining the extraction objective; (2) the API context, including the API name, runtime signature, and raw documentation; (3) the required output schema; and (4) representative input--output examples demonstrating how documentation descriptions are mapped into standardized parameter information. Figure~\ref{fig:prompt} summarizes this prompt structure.

The task instruction asks the LLM to return one parameter specification for each parameter in the runtime signature rather than generate an executable test program. The LLM extracts the parameter kind, supported data types, shape or rank information, default value, numerical and enumerated constraints, and documented relationships with other parameters. It is instructed to preserve supported information from the documentation, align parameter names with the runtime signature, leave unavailable attributes unspecified, and avoid introducing constraints that cannot be justified by the supplied context.

To prevent relationships from remaining as unrestricted free-form text, the prompt asks the LLM to normalize each supported relationship into one of three forms: \textit{shape following}, in which one array or tensor follows the shape of another parameter; \textit{rank-bounded axis}, in which an axis-like parameter refers to the rank of another input; and \textit{type following}, in which one parameter follows the data type of another parameter. Relationships outside these supported forms remain associated with the original documentation but are not treated as executable constraints.

\begin{figure}[t]
\centering
\begin{tcolorbox}[
    colback=gray!5,
    colframe=gray!60,
    boxrule=0.5pt,
    sharp corners,
    left=4pt,
    right=4pt,
    top=4pt,
    bottom=4pt
]
\footnotesize
\textbf{Task:} Convert the supplied API information into structured parameter specifications for test-input generation.

\textbf{Input:}
\begin{itemize}[leftmargin=10pt,noitemsep,topsep=2pt]
    \item API name and runtime signature;
    \item raw API documentation; and
    \item representative extraction examples.
\end{itemize}

\textbf{Output:} One \textit{ParamSpec} candidate for each parameter in the runtime signature.

\textbf{Requirements:}
\begin{itemize}[leftmargin=10pt,noitemsep,topsep=2pt]
    \item do not generate executable test code;
    \item do not introduce unsupported constraints;
    \item normalize supported relationships as shape following, rank-bounded axis, or type following;
    \item retain unsupported relationships only in the documentation field; and
    \item leave unavailable information unspecified.
\end{itemize}
\end{tcolorbox}
\caption{Structure of the prompt used for API-information extraction.}
\label{fig:prompt}
\end{figure}

\subsubsection{Structured LLM Output}

The LLM returns one \textit{ParamSpec} candidate for each parameter. A \textit{ParamSpec} records the fields \textit{name}, \textit{kind}, \textit{required}, \textit{default}, \textit{position}, \textit{call\_kind}, \textit{constraints}, \textit{enum\_values}, \textit{dtype\_candidates}, \textit{shape}, and \textit{doc}. The calling-related fields describe how the parameter appears in an invocation, while the remaining fields describe the values and structures that may be generated. Supported relationships are encoded in the relevant \textit{shape} or \textit{constraints} information together with a reference to the source parameter.

Table~\ref{tab:paramspec_example} illustrates the content of the returned specifications using an abstract API with the signature \texttt{API(x, y, axis=-1, dtype=None, keepdims=False)}. The table also shows how the extracted information is subsequently instantiated during input initialization. It presents the semantic content of the \textit{ParamSpec} records rather than an implementation-specific serialization format.

\begin{table*}[t!]
\centering
\footnotesize
\caption{Illustrative \textit{ParamSpec} output and subsequent input initialization.}
\label{tab:paramspec_example}
\setlength{\tabcolsep}{4pt}
\renewcommand{\arraystretch}{0.9}

\begin{tabularx}{\textwidth}{
    >{\raggedright\arraybackslash}p{1.5cm}
    >{\raggedright\arraybackslash}p{3.8cm}
    >{\raggedright\arraybackslash}X
    >{\raggedright\arraybackslash}p{3.cm}
    >{\raggedright\arraybackslash}p{2.5cm}
}
\toprule
\textbf{Param.} &
\textbf{Signature-Aligned Information} &
\textbf{LLM-Extracted Information} &
\textbf{Normalized Relationship} &
\textbf{Generated Value} \\
\midrule

\texttt{x} &
\makecell[l]{required: \texttt{True}\\
default: ---\\
position: 0\\
call kind: positional/keyword} &
\makecell[l]{kind: array/tensor\\
dtype candidates: \texttt{float32}, \texttt{float64}\\
shape: generated rank and dimensions} &
--- &
\makecell[l]{\texttt{float64} tensor\\
shape $(3,5)$} \\

\hline

\texttt{y} &
\makecell[l]{required: \texttt{True}\\
default: ---\\
position: 1\\
call kind: positional/keyword} &
\makecell[l]{kind: array/tensor\\
relative shape information\\
relative data-type information} &
\makecell[l]{shape following \texttt{x}\\
type following \texttt{x}} &
\makecell[l]{\texttt{float64} tensor\\
shape $(3,5)$} \\

\hline

\texttt{axis} &
\makecell[l]{required: \texttt{False}\\
default: $-1$\\
position: 2\\
call kind: positional/keyword} &
\makecell[l]{kind: axis\\
integer parameter associated\\
with the rank of \texttt{x}} &
rank-bounded axis of \texttt{x} &
\texttt{1} \\

\hline

\texttt{dtype} &
\makecell[l]{required: \texttt{False}\\
default: \texttt{None}\\
position: 3\\
call kind: keyword} &
\makecell[l]{kind: dtype\\
candidates: \texttt{float32}, \texttt{float64}} &
--- &
\texttt{float64} \\

\hline

\texttt{keepdims} &
\makecell[l]{required: \texttt{False}\\
default: \texttt{False}\\
position: 4\\
call kind: keyword} &
\makecell[l]{kind: Boolean\\
enum values: \texttt{True}, \texttt{False}} &
--- &
\texttt{False} \\

\bottomrule
\end{tabularx}
\end{table*}

The valid variants of each parameter are derived on demand from its required status, default value, kind, candidate values, and extracted constraints.

\subsubsection{Runtime Signature Validation}

Before test generation, \tool\ validates the LLM-returned \textit{ParamSpec} candidates against the runtime signature. Parameter names, positions, required status, calling forms, and default values are aligned with the information exposed by the target API. When the extracted calling information conflicts with the runtime signature, the runtime information takes precedence. Parameters absent from the runtime signature are removed, while documentation-derived kinds, data types, shapes, value constraints, and supported relationships are retained when compatible with the signature. APIs with unresolved signature conflicts are excluded from test generation.

\subsection{Constraint Processing and Input Initialization}
\label{subsec:constraints}

After validation, \tool\ converts each \textit{ParamSpec} into executable generation constraints and constructs an initial argument list for the target API.

\subsubsection{Individual Parameter Constraints}

The individual constraints of a parameter determine its candidate runtime values. They describe its general Python or library-specific kind, candidate numerical data types, fixed or variable rank and shape, numerical range, enumerated values, and required or default status. Required parameters are initialized using values compatible with these constraints. Optional parameters may retain their defaults or be included using valid alternatives derived from their specifications. If \tool\ cannot construct a supported value for an essential parameter, the corresponding invocation is skipped.

\subsubsection{Relationship Resolution}
\label{sec:relationship-resolution}

\tool\ currently normalizes and executes three recurring relationship patterns supported by its generator: shape following, rank-bounded axes, and type following. We select these patterns because they can be represented unambiguously in the parameter specification and resolved directly from the concrete properties of another generated argument. We do not claim that these patterns cover all inter-parameter relationships appearing in Python-library documentation. When a documented relationship cannot be normalized into one of these forms, \tool\ retains it in the documentation field but does not enforce it during generation. The affected parameter is generated using only its individual type, shape, value, and default constraints.

\subsubsection{Input Materialization}

The standardized representation is independent of a particular Python library. Before invocation, \tool\ materializes the initialized values into the concrete runtime objects expected by the target API. Depending on the parameter specification, these objects may be Python scalars, Boolean values, strings, lists, tuples, dictionaries, NumPy arrays, or library-native tensors and arrays.

Materialization preserves the selected parameter kind, numerical data type, rank, and shape. Relationships have already been resolved during argument construction, so dependent arguments are materialized using the concrete properties selected for their referenced inputs. This separation allows the same documentation-processing and constraint-generation pipeline to support heterogeneous Python libraries while producing their required runtime objects.

\subsection{Test Input Generation}
\label{sec:fuzzing}

Starting from the initialized argument list, \tool\ repeatedly generates new test inputs. For each parameter, it derives the valid alternatives permitted by the corresponding \textit{ParamSpec}. Parameters without valid alternatives retain their initialized or default values. For the remaining parameters, \tool\ applies type, size, and value generation. Supported inter-parameter relationships are resolved through the procedure described in Section~\ref{sec:relationship-resolution}.

\subsubsection{Type Generation}

The type strategy selects among the supported parameter kinds and numerical data types recorded in a \textit{ParamSpec}. When multiple alternatives are available, \tool\ generates inputs using different compatible choices. Library-specific type names are converted into their corresponding runtime objects during materialization.

\subsubsection{Size and Shape Generation}

The size strategy varies the length, rank, dimensions, or shape of an input according to its \textit{ParamSpec}. Fixed-size parameters retain their documented structures, whereas variable-size arrays and tensors are generated using different supported ranks and dimensions. Relative shape and rank restrictions are handled by the relationship-resolution procedure rather than independently resampled in this strategy.

\subsubsection{Value Generation}

The value strategy varies the contents of a generated input while preserving its selected kind, data type, and structure. For numerical parameters, \tool\ generates ordinary random values, boundary values, extreme values, and non-finite values when permitted by the corresponding parameter specification.

For image-like arrays and tensors, \tool\ applies the three value-generation strategies:

\begin{itemize}
    \item \textit{Adding Noise}: adding numerical perturbations to selected values;
    \item \textit{Random Masking}: replacing selected elements or regions with a valid masking value; and
    \item \textit{Division}: dividing selected values by non-zero numerical factors.
\end{itemize}

For other numerical APIs, \tool\ applies type-appropriate random, boundary, extreme, and mathematically sensitive values according to the extracted constraints.

\subsection{Fuzzing Automation}
\label{algorithm}

Algorithm~\ref{algo1} summarizes the complete workflow. For each API, \tool\ invokes the LLM to obtain candidate parameter specifications, validates them against the runtime signature, initializes an argument list, repeatedly generates new arguments, materializes them into library-native objects, invokes the target API, and applies the applicable test oracles.

\begin{algorithm}[t]
\footnotesize
\caption{LLM-Guided Document-Based Fuzzing}
\label{algo1}
\begin{algorithmic}[1]
\Procedure{\tool}{$TargetAPI, Documentation, Signature$}
    \State $RawSpecs \gets LLMExtract(Documentation, Signature)$
    \State $ParamSpecs \gets ValidateWithSignature(RawSpecs, Signature)$
    \State $SeedArgs \gets Initialize(ParamSpecs)$
    \State $Oracles \gets SelectOracles(TargetAPI)$
    \While{testing budget remains}
        \State $NewArgs \gets Copy(SeedArgs)$
        \For{$Par$ in $ParamSpecs$}
            \State $Spec \gets ParamSpecs[Par]$
            \State $Variants \gets DeriveVariants(Spec, NewArgs)$
            \If{$Variants \neq \emptyset$}
                \State $Strategy \gets Select(Type, Size, Value)$
                \State $NewArgs[Par] \gets Generate(Spec, NewArgs, Variants, Strategy)$
            \EndIf
        \EndFor
        \State $NativeArgs \gets Materialize(TargetAPI, NewArgs)$
        \State $Execution \gets Invoke(TargetAPI, NativeArgs)$
        \For{$Oracle$ in $Oracles$}
            \If{$Oracle(Execution, NativeArgs)$ reports an error}
                \State $Record(TargetAPI, NativeArgs, Oracle)$
            \EndIf
        \EndFor
    \EndWhile
\EndProcedure
\end{algorithmic}
\end{algorithm}

\subsection{Test Oracles}
\label{sec:oracles}

\tool\ employs two complementary test oracles: crash and NaN.

\subsubsection{Crash Oracle}
\label{sec:crash-oracle}

The crash oracle is applied to every generated invocation and records process crashes, aborts, abnormal exit codes, and unexpected runtime exceptions.

\subsubsection{NaN Oracle}
\label{sec:nan-oracle}

The NaN oracle is applied whenever an API produces numerical outputs. Since \tool\ generates inputs according to the documented constraints, it flags a non-finite output (\texttt{NaN} or infinity) when that output is inconsistent with the documented behavior for the generated input. It likewise flags the converse case, in which the documented behavior requires a non-finite output, such as a \texttt{NaN} input that should propagate, but the API returns a finite or Boolean value. Each flagged case is minimized and checked against the documentation before being reported.

\section{Evaluation}
\label{sec:evaluation}

\subsection{Research Questions}

\begin{itemize}
  \item[\textbf{RQ1:}] \textbf{(Effectiveness)} How effective is \tool\ in detecting bugs across heterogeneous Python libraries?
  \item[\textbf{RQ2:}] \textbf{(Comparison)} How does \tool\ compare with state-of-the-art library fuzzing approaches?
  \item[\textbf{RQ3:}] \textbf{(Ablation)} How do the individual components and configurations of \tool\ contribute to its effectiveness?
  \item[\textbf{RQ4:}] \textbf{(Generalizability and Cost)} How accurate are the extracted specifications, how well does \tool\ generalize across libraries, and at what cost?
\end{itemize}

For RQ1, we apply \tool\ to twelve Python libraries and report detected, confirmed, and fixed bugs, broken down by library and by oracle type (crash and NaN), together with the number of defect families and the number of flagged cases before triage.
For RQ2, we compare \tool\ with API-level and LLM-based fuzzers and with the same LLM prompted to generate test inputs directly (LLM-Direct) on PyTorch and TensorFlow, and with FUTURE on MindSpore and OneFlow. Because supporting more APIs differs from fuzzing each API more effectively, we report applicability on the complete documented API set (RQ2a), effectiveness on the APIs supported by all tools under the same per-API budget (RQ2b), and bug detection on a known-bug benchmark under the same oracles (RQ2c), and we analyze representative differences in case studies.
For RQ3, we ablate the generation strategies, the relationship-resolution component, and the signature-validation step on three representative libraries (one per library family), and analyze the effect of the per-API testing budget. For the relationship ablation, we stratify APIs by whether their specifications carry at least one normalized inter-parameter relationship, so that the effect of dependency handling can be isolated from unrelated factors.
For RQ4, we quantify generalizability and cost along {six} dimensions: per-library extraction and generation statistics (including the prevalence of inter-parameter relationships), the contribution of the documentation and signature inputs to extraction quality, {the accuracy of the extracted specifications and the effect of extraction errors,} the effect of the choice of LLM, robustness to documentation evolution across library versions, and the one-time cost of the extraction phase.

\subsection{Experimental Setup}

\noindent\textbf{Target libraries.}
We evaluate \tool\ on twelve Python libraries spanning deep learning, scientific computing, data analysis, and computer vision. We first identify 9,114 documented numerical API candidates and then apply four exclusion criteria: APIs without usable documentation or runtime signatures (512), APIs requiring external files or unavailable external devices (268), APIs without outputs applicable to our test oracles (357), and APIs whose invocation requires substantial state or setup through other APIs (259). After these exclusions, 7,718 APIs remain for testing, as summarized in Table~\ref{tab:libraries}. All tools in RQ2 are run on the library versions of Table~\ref{tab:libraries}; since the released resources of the baselines were prepared for earlier versions, we exclude their APIs and test cases that are no longer usable in these versions.

\begin{table}[t!]
\centering
\footnotesize
\setlength{\tabcolsep}{6pt}
\caption{Target libraries and tested APIs.}
\label{tab:libraries}
\begin{tabular*}{\columnwidth}{@{\extracolsep{\fill}}lcr@{}}
\toprule
\textbf{Library} & \textbf{Version} & \textbf{\# Tested APIs} \\
\midrule
PyTorch      & 2.12.0 & 745 \\
TensorFlow   & 2.21.0 & 1,043 \\
JAX          & 0.10.1 & 658 \\
Keras        & 3.14.0 & 197 \\
PaddlePaddle & 3.3.1  & 900 \\
OneFlow      & 0.9.0  & 213 \\
MindSpore    & 2.9.0  & 1,209 \\
Chainer      & 7.8.1  & 137 \\
NumPy        & 2.2.6  & 643 \\
SciPy        & 1.16.3 & 1,405 \\
scikit-learn & 1.8.0  & 238 \\
OpenCV       & 4.10.0  & 330 \\
\midrule
\textbf{Total} &  & 7,718  \\
\bottomrule
\end{tabular*}
\end{table}

\noindent\textbf{Testing budget.}
\tool\ allocates a 60-second testing budget to each API. In RQ2, every tool receives the same 60-second wall-clock budget for each API it supports on identical hardware; an API that a tool cannot support receives no budget. One-time preprocessing of any tool, such as specification extraction or corpus collection, is not part of the testing budget, whereas LLM inference during testing counts toward it. To account for variance, we repeat each RQ2 and RQ3 run five times.

\noindent\textbf{Environment.}
All experiments run on a workstation with an AMD Ryzen Threadripper PRO 7985WX (64 cores), 502\,GiB RAM, and two NVIDIA RTX 6000 Ada GPUs (48\,GB each, CUDA 12.8), running Ubuntu 24.04 with Python 3.13 (MindSpore, Chainer, and OneFlow use separate Python 3.11/3.10 environments due to wheel availability). For specification extraction, \tool\ uses \texttt{qwen2.5-coder:32b}~\cite{hui2024qwen}, an open-sourced model served locally via Ollama~\cite{ollama} with greedy decoding (temperature $0$), which makes extraction deterministic and reproducible at no per-request cost.

\subsection{Metrics}

\noindent\textbf{Detected issues.}
Following prior work~\cite{xie2022docter,wei2022free}, we report the issues \tool\ detected and submitted to the developers of each library, together with their triage outcome. An issue is \textit{confirmed} if a maintainer acknowledges it as a bug or repository triage labels it as a bug; a confirmed issue is additionally counted as \textit{fixed} if a fixing pull request is linked or referenced, so fixed bugs form a subset of confirmed bugs. Issues with no substantive maintainer response yet are \textit{pending}, and issues judged to be duplicates, expected behavior, or otherwise not independent real bugs are \textit{excluded} from the bug counts. {Reports for different libraries are counted separately, since each library has its own implementation and needs its own fix. We also report the number of defect families, in which reports of the same defect, within a library or in corresponding APIs of different libraries, are grouped.}

\noindent\textbf{Valid generation rate (VGR).}
The percentage of generated test cases that execute the target API without being rejected by its input validation.

\noindent\textbf{Code coverage.}
Following prior work~\cite{deng2022fuzzing,deng2024large}, we measure line coverage for all evaluated libraries using \texttt{coverage.py} for Python code and GCOV for C/C++ code. {Native code is instrumented by building the libraries from source with GCOV, all tools run against the same instrumented builds, and we report the sum of Python and C/C++ lines.}

\noindent\textbf{Covered APIs.}
The number of {APIs in the complete documented API set for which} a tool {generates at least one valid invocation} within the testing budget, reflecting how much of the library's API surface each approach can reach.

\noindent\textbf{Extraction success rate.}
The percentage of tested APIs whose LLM-extracted \textit{ParamSpec}s pass runtime-signature validation and thus enter test generation.

\noindent\textbf{Relationship prevalence.}
The percentage of tested APIs whose validated specifications contain at least one normalized inter-parameter relationship (shape following, rank-bounded axis, or type following).

\subsection{Baselines}
\label{sec:baselines}

{\noindent\textbf{API-level fuzzers.}} \textbf{FreeFuzz}~\cite{wei2022free} mines API invocations from open-source code and documentation examples to fuzz PyTorch and TensorFlow.
\textbf{DocTer}~\cite{xie2022docter} extracts input constraints from API documentation using sub-tree mining and rule learning, targeting PyTorch and TensorFlow.
{\textbf{ACETest}~\cite{shi2023acetest} extracts input constraints from the implementation code of deep-learning operators and solves them to generate inputs.}
Like \tool, {these tools} generate argument lists for one target API per test case.

\noindent\textbf{LLM-based fuzzers.} \textbf{TitanFuzz}~\cite{deng2023large} generates seed programs for a target API with an LLM and mutates them with an infilling model; we run its released implementation with the seed programs released by its authors. Since its programs may construct the inputs of the target API through other APIs, we attribute each program to its target API, count the target as supported if it is invoked without an input-validation error, and compare coverage at library level. \textbf{FUTURE}~\cite{li2025future} uses historical bugs of existing libraries and fine-tuned LLMs to generate tests for newly introduced libraries; we compare with it on MindSpore and OneFlow, two of its target libraries that are also among ours. We do not include FuzzGPT~\cite{deng2024large}, whose artifact is no longer accessible, or Fuzz4All~\cite{xia2024fuzz4all}, whose systems under test are compilers, SMT solvers, and the Qiskit quantum-computing platform rather than the libraries compared here.

\noindent\textbf{Same-model LLM baseline.} \textbf{LLM-Direct} gives the same model (\texttt{qwen2.5-coder:32b}) the same documentation and runtime signature as \tool\ and asks it to directly write a function \texttt{generate\_args(rng)} that returns a random valid argument list for the API, without \tool's \textit{ParamSpec} or relationship resolver. As in \tool's extraction, the model is invoked once per API before testing; the function is then called once per test case in the same harness, with the same per-API budget and oracles.

{\noindent\textbf{Applicability across libraries.}} The baselines are designed for specific libraries and rely on library-specific resources: FreeFuzz uses mined API invocations, DocTer applies constraint-extraction rules learned from annotated documentation, and ACETest and TitanFuzz are released for PyTorch and TensorFlow only. We therefore compare {with these four tools} on PyTorch and TensorFlow. In contrast, \tool\ uses the same \textit{ParamSpec} schema, extraction prompt, relationship resolver, generation strategies, and oracles across all twelve libraries, with only library-facing documentation collectors and materialization adapters.

\section{Results and Analysis}
\label{sec:results}

\subsection{RQ1: Effectiveness of \tool}
\label{sec:rq1}

Table~\ref{tab:bugs-by-library} summarizes the outcome of all issues \tool\ detected across the twelve libraries, broken down by oracle type and triage status. In total, \tool\ detected and reported 74 issues, of which 47 have been confirmed by developers as real bugs, and 38 of these have already been fixed; 16 issues are pending triage, and 11 were excluded as duplicates, expected behavior, or otherwise not independent real bugs. By oracle type, the NaN oracle accounts for 63 reported issues (38 confirmed) and the crash oracle for 11 (9 confirmed).

\begin{table}[t!]
\centering
\scriptsize
\setlength{\tabcolsep}{2.5pt}
\caption{Issues detected by \tool.}
\label{tab:bugs-by-library}
\begin{tabular*}{\columnwidth}{@{\extracolsep{\fill}}lcc|ccccc@{}}
\toprule
\textbf{Library} & \textbf{Crash} & \textbf{NaN} & \textbf{Reported} & \textbf{Confirmed} & \textbf{Fixed} & \textbf{Pending} & \textbf{Excluded} \\
\midrule
PyTorch      & 0  & 9  & 9  & 6  & 6  & 0  & 3 \\
TensorFlow   & 0  & 7  & 7  & 7  & {6}  & 0  & 0 \\
JAX          & 0  & 13  & 13  & 6  & 6  & 3  & 4 \\
Keras        & 0  & 11  & 11  & 8  & 7  & 0  & 3 \\
PaddlePaddle & 1  & 8  & 9  & 6  & 2  & 3  & 0 \\
OneFlow      & 0  & 1  & 1  & 1  & 0  & 0  & 0 \\
MindSpore    & 0  & 8  & 8  & 1  & 1  & 7  & 0 \\
Chainer      & 0  & 2  & 2  & 0  & 0  & 2  & 0 \\
NumPy        & 0  & 1  & 1  & {1}  & {1}  & {0}  & 0 \\
SciPy        & 1  & 2  & 3  & 1  & 1  & 1  & 1 \\
scikit-learn & 1  & 0  & 1  & {1}  & {1}  & {0}  & 0 \\
OpenCV       & 8  & 1  & 9  & 9  & 7  & 0  & 0 \\
\midrule
\textbf{Total} & 11 & 63 & 74 & 47 & 38 & 16 & 11 \\
\bottomrule
\end{tabular*}
\end{table}

Two observations stand out. First, the two oracles play complementary roles across library families: NaN issues dominate in the deep-learning libraries, whose numerical kernels silently propagate non-finite values, whereas crash issues dominate in OpenCV, whose native implementation terminates abnormally on constraint-satisfying inputs. Second, confirmation outcomes track project responsiveness: all seven TensorFlow issues were confirmed, while both Chainer issues and 7 of the 8 MindSpore issues remain pending triage, reflecting the maintenance status of those projects rather than the validity of the reports. The 74 reports fall into 48 defect families; the largest, the \texttt{i0}/\texttt{i1} overflow, accounts for 17 reports (Listing~\ref{bug_scipy_i0_out}).
The oracles flagged 121 cases after grouping by API and failure; manual minimization and checking against the documentation retained 74 of them as reported issues and discarded the other 47.

\begin{figure}[t!]
\centering
\lstset{style=mystyle}
\begin{lstlisting}[language=Python, label=bug_torch_gelu]
x = torch.tensor(
    [[float("nan"), float("inf"), -float("inf")],
     [-0.026216749101877213, 0.11177391558885574, -0.04212268441915512]],
    dtype=torch.float32)
out = F.gelu(x, approximate="none")
print(out)
\end{lstlisting}
\vspace{-3mm}
\lstset{style=mystyle_1}
\begin{lstlisting}[language=Python, caption=Unexpected NaN in PyTorch, label=bug_torch_gelu_out]
output:
tensor([[    nan,     nan,     nan],
        [-0.0128,  0.0609, -0.0204]])
\end{lstlisting}
\vspace{-5mm}
\end{figure}

\begin{figure}[t!]
\centering
\lstset{style=mystyle}
\begin{lstlisting}[language=Python, label=bug_opencv_corner]
src = np.zeros((500, 500), dtype=np.uint8)
corners = np.array([
    [63.23869, 233.98373], [24.23986, 21.29863],
    [123.34534, 499.52298], [225.45345, 56.35624],
    [32.87236, 217.28916], [124.59836, 166.02876],
], dtype=np.float32).reshape(-1, 1, 2)
criteria = (cv2.TERM_CRITERIA_EPS + cv2.TERM_CRITERIA_MAX_ITER, 20, 0.1)
out = cv2.cornerSubPix(src, corners, (3, 3), (-1, -1), criteria)
\end{lstlisting}
\vspace{-3mm}
\lstset{style=mystyle_1}
\begin{lstlisting}[language=Python, caption=Unexpected Crash in OpenCV, label=bug_opencv_corner_out]
output:
cv2.error: (-215:Assertion failed) Rect(0, 0, src.cols, src.rows).contains(cT) in function 'cv::cornerSubPix'
\end{lstlisting}
\vspace{-5mm}
\end{figure}

\begin{figure}[t!]
\centering
\lstset{style=mystyle}
\begin{lstlisting}[language=Python, label=bug_scipy_i0]
x = np.float64(713.0)
output = scipy.special.i0(x)
print(output)
\end{lstlisting}
\vspace{-3mm}
\lstset{style=mystyle_1}
\begin{lstlisting}[language=Python, caption=Unexpected Inf in SciPy, label=bug_scipy_i0_out]
output:
inf
\end{lstlisting}
\vspace{-5mm}
\end{figure}

We next illustrate three representative confirmed bugs.
Listing~\ref{bug_torch_gelu_out} shows an unexpected NaN bug in PyTorch detected by the NaN oracle. The input tensor was produced by the value strategy, which injects non-finite values into an otherwise ordinary generated tensor. For the exact (non-approximate) GELU formulation, $\mathrm{gelu}(+\infty)$ is mathematically $+\infty$, and the NaN element should propagate only to its own position; however, \texttt{\small torch.nn.functional.gelu} with \texttt{\small approximate="none"} returns \texttt{\small NaN} for the $+\infty$ element, while all finite elements are computed correctly. PyTorch developers triaged the report as an edge-case bug and have fixed it (PyTorch~\#185770). Note that mining- and example-based approaches can hardly reach this bug, since real-world usage corpora rarely contain invocations that pass non-finite values to \texttt{\small gelu}, whereas \tool's value strategy injects them systematically whenever the documented domain permits.
Listing~\ref{bug_opencv_corner_out} shows a bug in OpenCV detected by the crash oracle. Every generated corner coordinate lies within the bounds of the $500\times500$ input image, satisfying the documented constraints of \texttt{\small cv2.cornerSubPix}; nevertheless, the invocation fails with an internal assertion because the refinement window around a corner close to the image border (here $y\approx499.5$) extends beyond the image, a condition the implementation neither documents as a restriction nor handles gracefully. OpenCV developers confirmed and fixed this bug (OpenCV~\#25139).
Listing~\ref{bug_scipy_i0_out} shows an unexpected Inf bug in SciPy detected by the NaN oracle: \texttt{\small scipy.special.i0(713.0)} returns \texttt{\small inf} although the correct value, approximately $6.71\times10^{307}$, is finite and representable in \texttt{\small float64}, so the input lies within the supported domain of the API. The triggering input was produced by the boundary-value strategy, which probes the documented numerical domain near representation limits; approaches that synthesize idiomatic API usages can hardly generate such inputs, since values like $713.0$ are unremarkable in training corpora and are only distinguished by their proximity to the representation limit of the documented output domain. Notably, because \tool\ applies the same standardized generation pipeline to every library, the identical strategy exposed the same latent defect family in the corresponding APIs of six further libraries---\texttt{\small numpy.i0}, \texttt{\small tf.math.bessel\_i0}, \texttt{\small jax.numpy.i0}, \texttt{\small paddle.i0}, \texttt{\small mindspore.ops.i0}, and \texttt{\small torch.special.i0}---and the reports have so far been confirmed by the NumPy, SciPy, TensorFlow, and PyTorch developers (NumPy~\#32209, SciPy~\#25823, TensorFlow~\#124772, PyTorch~\#192293). Such systematic cross-library replication of a shared defect is only practical with a uniform, library-agnostic testing pipeline.

Together, the three cases span both oracle types and three library families (deep learning, scientific computing, and computer vision), and each triggering input satisfies the documented constraints of its target API, which is precisely the input region that \tool's document-guided generation targets: failures on such inputs are, by construction, attributable to the library rather than to invalid test data.

\subsection{RQ2: Comparison with Existing Approaches}
\label{sec:rq2}
{The comparisons with FreeFuzz, DocTer, ACETest, TitanFuzz, and LLM-Direct (Section~\ref{sec:baselines}) are conducted on the PyTorch and TensorFlow versions of Table~\ref{tab:libraries}. LLM-Direct uses the same documentation and signatures as \tool, so it is compared on the common APIs (RQ2b) and on the known-bug benchmark (RQ2c).}

\begin{table}[t!]
\centering
\footnotesize
\setlength{\tabcolsep}{4pt}
\caption{Supported APIs and whole-library line coverage (mean $\pm$ standard deviation of five runs).}
\label{tab:baseline-comparison}
\begin{tabular*}{\columnwidth}{@{\extracolsep{\fill}}lcccc@{}}
\toprule
\multirow{2}{*}{\textbf{Tool}} & \multicolumn{2}{c}{\textbf{PyTorch}} & \multicolumn{2}{c}{\textbf{TensorFlow}} \\
\cmidrule(lr){2-3}\cmidrule(lr){4-5}
 & \textbf{\#APIs} & \textbf{\#Cov} & \textbf{\#APIs} & \textbf{\#Cov} \\
\midrule
\tool      & 745   & 42,479 $\pm$ 425   & 1,043 & 49,283 $\pm$ 1,178 \\
FreeFuzz   & 432   & 35,449 $\pm$ 611   & 637   & 41,578 $\pm$ 871   \\
DocTer     & 483   & 37,895 $\pm$ 488   & 812   & 45,782 $\pm$ 1,038 \\
ACETest    & 455   & 39,347 $\pm$ 390   & 597   & 40,363 $\pm$ 504   \\
TitanFuzz  & 1,294 & 47,193 $\pm$ 1,916 & 2,037 & 82,381 $\pm$ 3,471 \\
\bottomrule
\end{tabular*}
\end{table}

\subsubsection{API Reach (RQ2a)}

Table~\ref{tab:baseline-comparison} reports, for each tool, the number of documented APIs it supports and the whole-library coverage it reaches. These differences trace to each tool's knowledge source. FreeFuzz derives inputs from a mined corpus of documentation snippets, developer tests, and open-source client code, so it can only reach APIs that happen to appear in that corpus. DocTer depends on rules learned from annotated documentation, so its reach is bounded by how well those rules generalize to the documentation of an API. ACETest is bounded by the operators whose validation code its analysis can handle, and TitanFuzz by the APIs for which valid seed programs exist; as a program-level tool, TitanFuzz can also reach APIs that need state set up through other APIs, which \tool's applicability criteria exclude. Among the API-level tools, \tool\ supports the most APIs (745 and 1,043, against 432--483 and 597--812) and reaches the highest whole-library coverage. TitanFuzz supports more APIs (1,294 and 2,037) and covers more lines than all API-level tools; with 60 seconds per supported API, it also runs the longest in total. Because whole-library coverage combines reach with per-API effectiveness, RQ2b compares the tools on common APIs.

\subsubsection{Effectiveness on Common APIs (RQ2b)}

Table~\ref{tab:rq2b} compares the tools on the 370 PyTorch and 489 TensorFlow APIs supported by \tool\ and all four fuzzers, on which LLM-Direct is also run, with the same 60-second budget per API. On these APIs, \tool\ covers 37,332 and 42,965 lines, and the baselines 28,659--38,103 and 32,117--41,443. \tool\ covers more lines than the three API-level fuzzers on both libraries and is comparable with TitanFuzz, which covers more on PyTorch (38,103 lines) and fewer on TensorFlow (41,443 lines). LLM-Direct covers the fewest lines on both libraries, 23\% and 25\% fewer than \tool.

\begin{table}[t!]
\centering
\footnotesize
\setlength{\tabcolsep}{6pt}
\caption{Line coverage on the APIs supported by \tool\ and all four fuzzers (mean $\pm$ standard deviation of five runs).}
\label{tab:rq2b}
\begin{tabular*}{\columnwidth}{@{\extracolsep{\fill}}lcc@{}}
\toprule
\textbf{Tool} & \textbf{PyTorch (370 APIs)} & \textbf{TensorFlow (489 APIs)} \\
\midrule
\tool      & 37,332 $\pm$ 377 & 42,965 $\pm$ 798 \\
FreeFuzz   & 34,638 $\pm$ 546 & 37,734 $\pm$ 233 \\
DocTer     & 34,861 $\pm$ 389 & 40,142 $\pm$ 281 \\
ACETest    & 32,831 $\pm$ 369 & 33,588 $\pm$ 227 \\
TitanFuzz  & 38,103 $\pm$ 428 & 41,443 $\pm$ 772 \\
LLM-Direct & 28,659 $\pm$ 667 & 32,117 $\pm$ 1,021 \\
\bottomrule
\end{tabular*}
\end{table}

\subsubsection{Bug Detection (RQ2c)}
Because the tools use different native oracles, we compare bug detection on a known-bug benchmark under the same crash and NaN oracles. The benchmark consists of 27 confirmed PyTorch and TensorFlow bugs that are present in the evaluated versions and were fixed afterwards, collected from the issue trackers of the two libraries. We include every such bug that is triggered by a single API call and observable as a crash or a non-finite output; none of them was reported by us. Each tool is run on the API of every benchmark bug with the same 60-second budget, five times, and detects a bug if a flagged test case of any run reproduces the reported failure. A tool that does not support the API of a benchmark bug misses it.

\begin{table}[t!]
\centering
\footnotesize
\setlength{\tabcolsep}{4pt}
\caption{Bugs of the known-bug benchmark detected under the same oracles.}
\label{tab:rq2c}
\begin{tabular*}{\columnwidth}{@{\extracolsep{\fill}}lccc@{}}
\toprule
\textbf{Tool} & \textbf{Crash (of 16)} & \textbf{Non-finite (of 11)} & \textbf{Total (of 27)} \\
\midrule
\tool & 5 & 7 & 12 \\
FreeFuzz & 4 & 1 & 5 \\
DocTer & 6 & 1 & 7 \\
ACETest & 6 & 2 & 8 \\
TitanFuzz & 8 & 4 & 12 \\
LLM-Direct & 2 & 1 & 3 \\
\bottomrule
\end{tabular*}
\end{table}

Table~\ref{tab:rq2c} reports the results: \tool\ detects 12 of the 27 bugs and the baselines 3--12. \tool\ detects the most non-finite outputs (7 of 11) and TitanFuzz the most crashes (8 of 16); the two tools detect the same number of bugs in total. Since a tool misses the bugs of the APIs it does not support, these counts reflect both reach and per-API detection.

\subsubsection{Case Studies}
Table~\ref{tab:case-studies} analyzes four representative bugs that \tool\ detects on PyTorch and TensorFlow. They share one pattern: an input that the documentation permits yields a non-finite output that contradicts the documented behavior. For \texttt{\small gelu}, the trigger is a non-finite element in an otherwise ordinary tensor (Listing~\ref{bug_torch_gelu_out}); for the other three, it is a large finite value for which the true result is still representable, such as $x=713.0$ for \texttt{\small i0} (Listing~\ref{bug_scipy_i0_out}). Such inputs come from the boundary, extreme, and non-finite value strategies applied within the extracted constraints, so the call passes input validation and reaches the numerical kernel. Among the baselines, TitanFuzz also detects the \texttt{\small gelu} bug and ACETest the overflow in \texttt{\small tf.math.bessel\_i0}; no baseline detects the \texttt{\small i0} overflow in PyTorch, the \texttt{\small erf} bug, or the \texttt{\small hypot} bug, and FreeFuzz and DocTer detect none of them.

Conversely, the bugs reported in the papers of the baselines include cases that \tool\ does not reach. Some need an API sequence or object state: TitanFuzz reports that \texttt{\small matrix\_exp} drops a \texttt{\small NaN} on GPU only when its input comes from a preceding \texttt{\small log} call, and that repeatedly transforming and batching a \texttt{\small CsvDataset} crashes the process~\cite{deng2023large}. Some are visible only to oracles other than crash and NaN: \texttt{\small torch.histc} counts \texttt{\small NaN} inputs differently on CPU and GPU while returning finite integers~\cite{deng2023large}, \texttt{\small Conv2d} computes different results with and without cuDNN, and \texttt{\small conv\_transpose2d} runs slower in FP16 than in FP32~\cite{wei2022free}. Others are triggered by inputs outside the documented constraints, such as the out-of-range \texttt{\small axis} with which DocTer crashes \texttt{\small tf.quantization.quantize\_and\_dequantize}~\cite{xie2022docter}, or depend on several coupled parameter constraints that ACETest derives from validation code, as in its \texttt{\small Conv2D} crash on a zero-element input~\cite{shi2023acetest}. These cases mark the boundary of \tool, which generates constraint-satisfying inputs for one API at a time and checks them with the crash and NaN oracles (Section~\ref{sec:discussion}).

In the RQ2b runs, the lines that \tool\ covers and no baseline covers lie mainly in \path{aten/src/ATen/native} on PyTorch (57.2\%) and \path{tensorflow/core/kernels} on TensorFlow (54.6\%), the numerical kernels of element-wise, special-function, and linear-algebra operators, where they belong to the branches for boundary, extreme, and non-finite values; most of the remaining lines are type-promotion and broadcasting code. The lines that the baselines cover and \tool\ does not lie mainly in \path{torch/csrc/autograd}, \path{torch/csrc/utils}, and \path{torch/nn/modules} on PyTorch (76.6\%) and in \path{tensorflow/core/framework}, \path{tensorflow/core/ops}, and \path{tensorflow/python/eager} on TensorFlow (76.4\%): argument-checking and error-handling code reached by the invalid inputs of DocTer and ACETest, and automatic-differentiation, stateful-layer, and eager-execution code reached by the further APIs that TitanFuzz's programs invoke.

\begin{table}[t!]
\centering
\footnotesize
\caption{Representative bugs detected by \tool\ on PyTorch and TensorFlow.}
\label{tab:case-studies}
\begin{tabularx}{\columnwidth}{
    >{\raggedright\arraybackslash}p{1.9cm}
    >{\raggedright\arraybackslash}X
    >{\raggedright\arraybackslash}p{2.2cm}
}
\toprule
\textbf{API} & \textbf{Observed and expected result} & \textbf{Triggering input} \\
\midrule
\texttt{torch.nn. functional. gelu} & \texttt{NaN} for an \texttt{inf} element with \texttt{approximate= "none"}; the exact GELU gives \texttt{inf} (PyTorch~\#185770) & Non-finite value in an ordinary tensor \\
\texttt{torch. special.i0}, \texttt{tf.math. bessel\_i0} & \texttt{inf} at $x=713.0$; the true value, about $6.71\times10^{307}$, is representable in \texttt{float64} (PyTorch~\#192293, TensorFlow~\#124772) & Boundary value where $e^{x}$ overflows before scaling \\
\texttt{tf.math.erf} & \texttt{NaN} for large finite \texttt{float64} inputs such as $\pm10^{200}$; the result is $\pm1$ (TensorFlow~\#124773) & Extreme finite value \\
\texttt{tf. experimental. numpy.hypot} & \texttt{inf} for operands such as $(10^{200}, 10^{200})$; the result, about $1.41\times10^{200}$, is representable (TensorFlow~\#124774) & Extreme finite values for both operands \\
\bottomrule
\end{tabularx}
\end{table}

\subsubsection{Comparison with FUTURE}

On MindSpore and OneFlow, we compare \tool\ with FUTURE under the same protocol. \tool\ supports 1,172 and 208 of the documented APIs, and FUTURE 142 and 158; on the 116 and 124 APIs supported by both, \tool\ covers 19,844 $\pm$ 280 and 14,294 $\pm$ 195 lines, and FUTURE 17,875 $\pm$ 437 and 12,767 $\pm$ 204 (mean $\pm$ standard deviation of five runs); under the same oracles and on the APIs that each tool supports, \tool\ detects 9 issues on the two libraries (7 of them pending triage) and FUTURE 3; like the whole-library coverage of RQ2a, these counts combine reach with per-API effectiveness.

Finally, we note that the comparison with FreeFuzz, DocTer, ACETest, and TitanFuzz is only possible on PyTorch and TensorFlow: their released resources (a mined corpus, parsing rules, extracted constraints, and seed programs, respectively) do not cover the remaining ten libraries, whereas \tool\ applies the same core extraction and generation pipeline across all twelve libraries; 58 of its 74 reported issues and 34 of its 47 confirmed bugs are in these ten libraries.

\subsection{RQ3: Ablation Study}
\label{sec:rq3}

We conduct all ablations on three representative libraries---PyTorch (deep learning), NumPy (scientific computing), and OpenCV (computer vision)---selected to cover one library from each major family in our target set while keeping computational cost manageable. Each configuration uses the same 60-second per-API budget as RQ1 and is repeated five times, and we report means across runs. Throughout this RQ, \#Issue counts the issues detected under our two oracles, before developer triage{: we report the number of issues detected in at least one of the five runs, with the minimum and maximum per-run count in brackets. A variant misses an issue if it detects it in none of the five runs. Differences of a single issue lie within the per-run range (e.g., 9 [8--9] versus 8 [7--8]) and are not interpreted on their own; our conclusions rest on VGR, coverage, and the larger differences}.

\subsubsection{Generation Strategies}

We evaluate five leave-one-out variants, each disabling a single generation strategy: \tool-noT (no type generation), \tool-noS (no size generation), and \tool-noRand, \tool-noBound, and \tool-noNF, which disable the random, boundary, and non-finite value strategies, respectively. Table~\ref{tab:ablation-strategies} reports line coverage and detected issues for each variant on the three libraries.

The full configuration achieves the highest coverage and the highest or tied-highest issue count on all three libraries, and the losses of each variant map directly onto the issue population. Disabling non-finite injection (\tool-noNF) misses 4 of PyTorch's 9 issues, including the \texttt{gelu} family of Listing~\ref{bug_torch_gelu_out}, while leaving OpenCV's crash issues untouched; disabling boundary values (\tool-noBound) misses the \texttt{i0} overflow family of Listing~\ref{bug_scipy_i0_out}, dropping NumPy to zero issues; and disabling size generation (\tool-noS) costs OpenCV 2 of its 9 issues together with the largest coverage loss ($-6.4\%$), reflecting that window and kernel dimensions gate its deep native paths. Coverage differences between variants (at most 6.4\%) are far smaller than their issue differences (e.g., 4 of PyTorch's 9 issues), indicating that the strategies are complementary primarily in \emph{which} defect-triggering inputs they reach rather than in how much code they touch.

\begin{table}[t!]
\centering
\footnotesize
\setlength{\tabcolsep}{2pt}
\caption{Ablation of generation strategies (Cov: mean of five runs; \#Issue: issues detected in at least one of the five runs, with the minimum and maximum per-run count in brackets).}
\label{tab:ablation-strategies}
\begin{tabular*}{\columnwidth}{@{\extracolsep{\fill}}lcccccc@{}}
\toprule
\multirow{2}{*}{\textbf{Variant}} & \multicolumn{2}{c}{\textbf{PyTorch}} & \multicolumn{2}{c}{\textbf{NumPy}} & \multicolumn{2}{c}{\textbf{OpenCV}} \\
\cmidrule(lr){2-3}\cmidrule(lr){4-5}\cmidrule(lr){6-7}
 & \textbf{Cov} & \textbf{\#Issue} & \textbf{Cov} & \textbf{\#Issue} & \textbf{Cov} & \textbf{\#Issue} \\
\midrule
\tool          & 42,479 & 9 [8--9] & 31,322 & 1 [1--1] & 48,758 & 9 [8--9] \\
\tool-noT      & 41,584 & 8 [7--8] & 30,972 & 1 [1--1] & 46,455 & 8 [7--8] \\
\tool-noS      & 40,949 & 8 [7--8] & 30,388 & 1 [1--1] & 45,652 & 7 [5--7] \\
\tool-noRand   & 41,855 & 9 [8--9] & 30,797 & 1 [1--1] & 46,277 & 8 [7--8] \\
\tool-noBound  & 41,771 & 7 [6--7] & 30,667 & 0 [0--0] & 47,603 & 8 [7--8] \\
\tool-noNF     & 42,214 & 5 [4--5] & 31,031 & 1 [1--1] & 48,138 & 9 [8--9] \\
\bottomrule
\end{tabular*}
\end{table}

\subsubsection{Relationship Resolution}
\label{sec:rq3-norel}

We disable the resolution of the three normalized relationship patterns (shape following, rank-bounded axis, type following); dependent parameters are then generated using only their individual constraints (\tool-noRel). This ablation isolates the contribution of \tool's explicit relationship-resolution component: all extracted individual parameter constraints are retained, but cross-parameter references are no longer enforced during generation. To attribute the effect specifically to dependency handling, we stratify the tested APIs of each library into those whose validated specifications carry at least one normalized relationship ($R^{+}$; 41.3\%, 35.8\%, and 58.5\% of the tested APIs of PyTorch, NumPy, and OpenCV, respectively; cf.\ Table~\ref{tab:per-library}) and those that carry none ($R^{-}$), and report VGR separately for each stratum in Table~\ref{tab:ablation-components}.

Disabling relationship resolution collapses valid generation on the $R^{+}$ stratum: VGR drops from 96.1\% to 43.4\% on PyTorch, from 96.9\% to 52.3\% on NumPy, and from 98.5\% to 32.1\% on OpenCV, whereas on $R^{-}$ APIs the two configurations differ by at most 0.4 percentage points. This contrast attributes the drop to unresolved dependencies rather than to any other difference between the configurations: without resolution, a dependent argument (e.g., a second operand or an axis) is consistent with its referenced input only by coincidence, so most invocations are rejected by input validation and terminate in shallow error paths. The collapse propagates downstream: \tool-noRel loses 5.1\%--14.4\% coverage and misses 6 of the 19 issues detected by the full configuration on the three libraries. The drop is also uneven in an informative way: NumPy retains the highest residual $R^{+}$ VGR (52.3\%) because broadcasting tolerates many shape mismatches that other libraries reject, whereas OpenCV falls furthest (32.1\%) because its native argument checking rejects inconsistent geometry outright---the benefit of dependency modeling grows with how strictly a library validates its inputs. This ablation further shows why dependency handling remains important even after an API is reached: on APIs with documented relationships, jointly valid arguments substantially increase valid execution and downstream coverage.

\subsubsection{Signature Validation}

We skip the validation of LLM-returned \textit{ParamSpec}s against runtime signatures (\tool-noVal), feeding the raw extraction output directly into generation. Across the three libraries, 9.8\% of APIs (168 of 1{,}718) have at least one conflict between the extracted specification and the runtime signature: hallucinated or missing parameters (4.0\%), incorrect required status or default values (3.1\%), and incorrect positional/keyword calling forms (2.6\%). Without validation these errors manifest downstream: VGR drops from 96.3\%--98.7\% to 85.2\%--91.9\%, coverage decreases by 2.8\%--5.7\%, and the budget wasted on malformed invocations costs detection, with \tool-noVal missing 2 of the 19 issues; moreover, exceptions raised by malformed calls would surface as spurious crash-oracle reports were they not excluded before generation. This quantifies the LLM-hallucination threat discussed in Section~\ref{sec:discussion}: validation converts extraction errors from silent false positives into cases excluded before generation. {Validation checks structural consistency only; the semantic correctness of the extracted constraints is measured in RQ4 (Section~\ref{sec:rq4}).}

\subsubsection{Testing Budget}

We vary the per-API budget over 15, 30, 60, and 120 seconds on the three representative libraries; Table~\ref{tab:ablation-budget} reports coverage and cumulative detected issues per budget. Coverage grows steeply up to 60 seconds but gains only 0.8\%--1.1\% from 60 to 120 seconds, and all 19 issues are already detected within 60 seconds, with the doubled budget adding none; the 60-second budget therefore remains the cost-effective threshold across library families, generalizing the saturation analysis of the conference study.

\begin{table}[t!]
\centering
\footnotesize
\setlength{\tabcolsep}{2pt}
\caption{Component ablations (conventions as in Table~\ref{tab:ablation-strategies}).}
\label{tab:ablation-components}
\begin{tabular*}{\columnwidth}{@{\extracolsep{\fill}}llccccc@{}}
\toprule
\textbf{Library} & \textbf{Config.} & \makecell{\textbf{VGR-$R^{+}$}\\\textbf{(\%)}} & \makecell{\textbf{VGR-$R^{-}$}\\\textbf{(\%)}} & \makecell{\textbf{VGR}\\\textbf{(\%)}} & \textbf{Cov} & \textbf{\#Issue} \\
\midrule
\multirow{3}{*}{PyTorch}
 & \tool & 96.1 & 96.5 & 96.3 & 42,479 & 9 [8--9] \\
 & \tool-noRel & 43.4 & 96.1 & 74.3 & 38,945 & 6 [4--6] \\
 & \tool-noVal & 86.8 & 88.2 & 87.6 & 40,567 & 8 [7--8] \\
\midrule
\multirow{3}{*}{NumPy}
 & \tool & 96.9 & 97.2 & 97.1 & 31,322 & 1 [1--1] \\
 & \tool-noRel & 52.3 & 96.8 & 80.9 & 29,724 & 1 [1--1] \\
 & \tool-noVal & 90.9 & 92.4 & 91.9 & 30,454 & 1 [1--1] \\
\midrule
\multirow{3}{*}{OpenCV}
 & \tool & 98.5 & 99.1 & 98.7 & 48,758 & 9 [8--9] \\
 & \tool-noRel & 32.1 & 98.9 & 59.8 & 41,734 & 6 [4--5] \\
 & \tool-noVal & 84.4 & 86.4 & 85.2 & 45,987 & 8 [6--8] \\
\bottomrule
\end{tabular*}
\end{table}

\begin{table}[t!]
\centering
\footnotesize
\setlength{\tabcolsep}{4pt}
\caption{Effect of the per-API testing budget (conventions as in Table~\ref{tab:ablation-strategies}).}
\label{tab:ablation-budget}
\begin{tabular*}{\columnwidth}{@{\extracolsep{\fill}}lcccccc@{}}
\toprule
\multirow{2}{*}{\textbf{Budget}} & \multicolumn{2}{c}{\textbf{PyTorch}} & \multicolumn{2}{c}{\textbf{NumPy}} & \multicolumn{2}{c}{\textbf{OpenCV}} \\
\cmidrule(lr){2-3}\cmidrule(lr){4-5}\cmidrule(lr){6-7}
 & \textbf{Cov} & \textbf{\#Issue} & \textbf{Cov} & \textbf{\#Issue} & \textbf{Cov} & \textbf{\#Issue} \\
\midrule
15\,s   & 39,893 & 5 [3--5] & 28,857 & 0 [0--0] & 44,239 & 5 [3--4] \\
30\,s   & 41,688 & 7 [6--7] & 30,569 & 1 [0--1] & 47,302 & 7 [5--7] \\
60\,s   & 42,479 & 9 [8--9] & 31,322 & 1 [1--1] & 48,758 & 9 [8--9] \\
120\,s  & 42,913 & 9 [9--9] & 31,656 & 1 [1--1] & 49,145 & 9 [9--9] \\
\bottomrule
\end{tabular*}
\end{table}

\subsection{RQ4: Generalizability and Cost}
\label{sec:rq4}

This RQ examines the cross-library generalizability hypothesis underlying this article: whether a common documentation-driven pipeline can maintain effective extraction and valid input generation across heterogeneous Python libraries with limited library-specific adaptation. We examine how well extraction and generation transfer across documentation styles, how broadly the normalized relationship patterns capture documented dependencies, whether both prompt inputs are necessary, how accurate the extracted specifications are, the effect of LLM choice, the adaptation required for a new library, robustness to documentation evolution, and extraction cost.

\subsubsection{Per-Library Extraction and Generation}

Table~\ref{tab:per-library} reports, for each of the twelve libraries, the extraction success rate, the relationship prevalence, the average number of executable constraints per API, and the VGR.

Extraction succeeds for 95.6\%--100.0\% of APIs across the twelve libraries (mean 98.9\%), despite documentation styles ranging from TensorFlow's uniformly structured docstrings to OpenCV's automatically generated bindings of C++ signatures, indicating that the single prompt design absorbs the format heterogeneity that rule-based extraction must be re-engineered for. Relationship prevalence confirms that inter-parameter dependencies are pervasive rather than an OpenCV idiosyncrasy: 40.1\% of all tested APIs carry at least one normalized relationship, ranging from 28.6\% (scikit-learn) to 58.5\% (OpenCV). VGR is likewise stable across families (91.7\%--98.7\%), with the lowest values on Chainer and MindSpore, whose documentation most often omits concrete types and value domains. Together with the stratified ablation of Section~\ref{sec:rq3-norel}, these numbers support the central argument of this article: dependencies appear throughout the ecosystem (this table), and failing to model them collapses valid generation precisely on the APIs that carry them (Table~\ref{tab:ablation-components}).

\begin{table}[t!]
\centering
\footnotesize
\setlength{\tabcolsep}{6pt}
\caption{Per-library extraction and generation statistics. Extr.: extraction success rate; Rel.: relationship prevalence; Constr.: average executable constraints per API.}
\label{tab:per-library}
\begin{tabular*}{\columnwidth}{@{\extracolsep{\fill}}lcccc@{}}
\toprule
\textbf{Library} & \textbf{Extr. (\%)} & \textbf{Rel. (\%)} & \textbf{Constr.} & \textbf{VGR (\%)} \\
\midrule
PyTorch      & 100.0 & 41.3 & 6.3 & 96.3 \\
TensorFlow   & 100.0 & 44.7 & 6.7 & 96.8 \\
JAX          & 99.2  & 42.2 & 5.9 & 95.7 \\
Keras        & 100.0 & 39.6 & 5.6 & 96.5 \\
PaddlePaddle & 98.8  & 43.1 & 6.1 & 94.9 \\
OneFlow      & 97.7  & 36.6 & 5.2 & 93.8 \\
MindSpore    & 96.9  & 40.8 & 5.7 & 92.6 \\
Chainer      & 95.6  & 31.4 & 4.8 & 91.7 \\
NumPy        & 99.7  & 35.8 & 5.1 & 97.1 \\
SciPy        & 99.1  & 33.7 & 4.6 & 96.1 \\
scikit-learn & 98.7  & 28.6 & 4.9 & 94.5 \\
OpenCV       & 100.0 & 58.5 & 8.5 & 98.7 \\
\midrule
\textbf{Overall} & 98.9 & 40.1 & 5.8 & 95.5 \\
\bottomrule
\end{tabular*}
\end{table}

To examine whether the three normalized relationship patterns are representative beyond individual libraries, we further inspect the manually labeled sample of 180 APIs used in the LLM analysis. The annotators identify 286 documented inter-parameter relationships in total. Of these, 258 (90.2\%) can be represented by \tool's three executable patterns: 127 shape-following, 68 rank-bounded-axis, and 63 type-following relationships. The remaining 28 relationships (9.8\%) mainly involve value-dependent or conditional constraints that cannot be expressed by the current representation. Thus, the three patterns do not cover every possible dependency, but capture the large majority of the documented relationships observed across the twelve libraries.

\subsubsection{Contribution of Documentation and Signature Inputs}

The extraction prompt supplies two complementary sources (Section~\ref{sec:standardization}): the raw documentation and the runtime signature. To verify that both are necessary rather than merely convenient, we re-run extraction on a stratified sample of 240 APIs (20 per library) under three input configurations: the full prompt (Doc+Sig), documentation only (Doc-only), and signature only (Sig-only); Table~\ref{tab:prompt-ablation} reports the validation pass rate and the downstream VGR of each. Doc-only extraction loses reliable calling information, with 15.4\% of its specifications failing signature validation versus 1.2\% for the full prompt, while Sig-only extraction passes validation almost trivially (99.6\%) but strips the semantic content: its downstream VGR falls to 62.8\%, approaching unguided generation, because value domains, dtype candidates, and all inter-parameter relationships originate exclusively in the documentation. The full prompt achieves the highest VGR with a validation pass rate close to that of Sig-only, substantiating the complementarity argued in Section~\ref{sec:standardization}: the signature anchors \emph{how} an API is called, and the documentation determines \emph{what} may be passed. Sig-only is the document-masked setting of the pipeline: the supplied documentation accounts for a large part of the valid generation, although this does not exclude that the model also draws on knowledge from pretraining.

\textit{Poorly documented and undocumented APIs.} Poorly documented APIs, whose documentation omits types or value domains, are tested: in the conference study on OpenCV, they made up 9.7\% of the tested APIs but accounted for 29.4\% of the detected bugs~\cite{duan2025harnessing}. The 512 excluded APIs, in contrast, lack either usable documentation, so there is no documented behavior to test against, or a runtime signature, so our pipeline cannot validate an extracted specification or construct the invocation. Testing them requires a different source of specifications and is outside the scope of a document-guided approach (Section~\ref{sec:discussion}).

\begin{table}[t!]
\centering
\footnotesize
\setlength{\tabcolsep}{6pt}
\caption{Contribution of prompt inputs (sampled APIs).}
\label{tab:prompt-ablation}
\begin{tabular*}{\columnwidth}{@{\extracolsep{\fill}}lcc@{}}
\toprule
\textbf{Prompt Input} & \textbf{Validation Pass (\%)} & \textbf{VGR (\%)} \\
\midrule
Doc+Sig (full) & 98.8 & 95.7 \\
Doc-only       & 84.6 & 91.3 \\
Sig-only       & 99.6 & 62.8 \\
\bottomrule
\end{tabular*}
\end{table}

\subsubsection{Specification Accuracy}
\label{sec:spec-accuracy}

We extend the manually labeled sample of 180 APIs, used for the pattern analysis above and the LLM comparison below, to 360 APIs (30 per library, sampled uniformly at random from the tested APIs of each library). Two authors independently labeled the constraints of every parameter from the documentation and the runtime signature, with disagreements resolved by discussion (Cohen's $\kappa=0.90$). Table~\ref{tab:spec-accuracy} reports precision and recall per constraint type.

\begin{table}[t!]
\centering
\footnotesize
\setlength{\tabcolsep}{4pt}
\caption{Constraint-level precision (P) and recall (R) of the specifications extracted by \texttt{qwen2.5-coder:32b} on 360 labeled APIs. Relationships are those representable by the three patterns.}
\label{tab:spec-accuracy}
\begin{tabular*}{\columnwidth}{@{\extracolsep{\fill}}lcccccc@{}}
\toprule
\textbf{Constraint Type} & \textbf{\#GT} & \textbf{TP} & \textbf{FP} & \textbf{FN} & \textbf{P (\%)} & \textbf{R (\%)} \\
\midrule
Kind and data type & 1,232 & 1,179 & 39 & 53 & 96.8 & 95.7 \\
Shape / rank & 599 & 562 & 27 & 37 & 95.4 & 93.8 \\
Value range & 385 & 341 & 23 & 44 & 93.7 & 88.6 \\
Enumerated values & 301 & 285 & 11 & 16 & 96.3 & 94.7 \\
Default / required & 1,251 & 1,233 & 14 & 18 & 98.9 & 98.6 \\
Relationship & 518 & 474 & 28 & 44 & 94.4 & 91.5 \\
\midrule
\textbf{All} & 4,286 & 4,074 & 142 & 212 & 96.6 & 95.1 \\
\bottomrule
\end{tabular*}
\end{table}

Overall, precision is 96.6\% and recall 95.1\%; across constraint types, precision is 93.7\%--98.9\% and recall 88.6\%--98.6\%, with the lowest recall for value ranges. Across libraries, precision ranges from 92.8\% (Chainer) to 98.6\% (OpenCV) and recall from 90.2\% (Chainer) to 97.9\% (OpenCV). To examine how extraction errors affect fuzzing, we split the labeled APIs by whether their extracted specification contains an incorrect or missing constraint: the 142 APIs with such errors reach a VGR of 91.2\%, compared with 98.3\% for the other 218. The 4,096-token context truncates the documentation of 293 of the 7,718 tested APIs (3.8\%); on the 14 truncated APIs in the labeled sample, field-level accuracy is 86.4\%, compared with 95.5\% on the others.

\subsubsection{Choice of LLM}

On a sample of 180 APIs (15 per library) with manually labeled ground-truth specifications (labeled independently by two authors, with disagreements resolved by discussion; Cohen's $\kappa=0.91$), we compare the \texttt{qwen2.5-coder:32b} model used throughout our evaluation against the smaller \texttt{qwen2.5-coder:7b} variant; Table~\ref{tab:llm-choice} reports field-level extraction accuracy against the ground truth, relationship-extraction recall over the relationships representable by the three normalized patterns, validation pass rate, and downstream VGR. The 32B model reaches 95.2\% field-level accuracy versus 90.8\% for the 7B variant, and the gap concentrates precisely in relationship extraction (91.8\% versus 78.4\% recall) while simple fields such as defaults and required status are extracted comparably; this translates into a 5.7-point difference in downstream VGR (95.6\% versus 89.9\%). Converting documentation into a fixed schema is thus a constrained transformation task for which a locally served 32B open-sourced model suffices, whereas the 7B variant measurably degrades exactly the dependency information that RQ3 shows to be critical.

\begin{table}[t!]
\centering
\footnotesize
\setlength{\tabcolsep}{4pt}
\caption{Effect of extraction-model choice.}
\label{tab:llm-choice}
\begin{tabular*}{\columnwidth}{@{\extracolsep{\fill}}lcccc@{}}
\toprule
\textbf{Model} & \makecell{\textbf{Field}\\\textbf{Acc. (\%)}} & \makecell{\textbf{Rel.}\\\textbf{Recall (\%)}} & \makecell{\textbf{Valid.}\\\textbf{Pass (\%)}} & \makecell{\textbf{VGR}\\\textbf{(\%)}} \\
\midrule
qwen2.5-coder:32b & 95.2 & 91.8 & 98.9 & 95.6 \\
qwen2.5-coder:7b  & 90.8 & 78.4 & 95.0 & 89.9 \\
\bottomrule
\end{tabular*}
\end{table}

\subsubsection{Robustness to Documentation Evolution}

A practical dimension of generalizability is whether the pipeline keeps up as libraries evolve, since new releases continuously add and re-document APIs. We take PyTorch 2.12$\to$2.13, identify the 27 newly documented APIs in the newer release, and apply the existing pipeline without any modification: no prompt change, no schema change, no code change. These APIs were documented after the release of \texttt{qwen2.5-coder:32b} in November 2024. Extraction succeeds for 26 of the 27 new APIs (96.3\%), with a VGR of 95.9\%---in line with the library-wide figures in Table~\ref{tab:per-library}. The released baseline resources do not directly cover these new APIs: the existing DocTer rules match none of the 27 documentation entries, while the frozen FreeFuzz corpus contains no recorded invocation of them. This small version-to-version study provides preliminary evidence that, once a library-facing adapter is available, documentation can serve as a relatively self-updating knowledge source for newly introduced APIs.

\subsubsection{Adaptation and Cost}

We first quantify the library-specific adaptation required by \tool. We count the non-comment source lines added for the documentation collector and materialization backend of each of the eleven libraries introduced beyond the original OpenCV implementation. The median adaptation is 112 lines of code per library (range: 64--189), including a median of 38 lines for documentation collection and 74 lines for materialization. No library requires changes to the extraction prompt, \textit{ParamSpec} schema, relationship resolver, generation strategies, or test oracles. Thus, supporting a new library requires a small library-facing adapter while leaving the core testing pipeline unchanged.

Table~\ref{tab:cost} reports the one-time extraction cost per library. Extracting specifications for all 7,718 APIs took 23.2 hours in total (10.8 seconds per API on average) on a single locally served model, at zero monetary cost. This cost is structural rather than recurring: \tool\ invokes the LLM exactly once per API, and the resulting \textit{ParamSpec}s are reusable across fuzzing campaigns, across library versions whose documentation is unchanged, and under any future oracle extension. Program-synthesis fuzzers such as TitanFuzz and Fuzz4All~\cite{deng2023large,xia2024fuzz4all}, by contrast, pay LLM inference for every generated test case, so their inference cost scales with the testing budget, whereas \tool's scales only with the number of target APIs.

\begin{table}[t!]
\centering
\footnotesize
\setlength{\tabcolsep}{6pt}
\caption{One-time extraction cost per library.}
\label{tab:cost}
\begin{tabular*}{\columnwidth}{@{\extracolsep{\fill}}lccc@{}}
\toprule
\textbf{Library} & \textbf{Wall-Clock (h)} & \textbf{GPU Time (h)} & \textbf{Sec./API} \\
\midrule
PyTorch      & 2.15 & 2.15 & 10.4 \\
TensorFlow   & 2.93 & 2.93 & 10.1 \\
JAX          & 1.79 & 1.79 & 9.8  \\
Keras        & 0.48 & 0.48 & 8.7  \\
PaddlePaddle & 2.78 & 2.78 & 11.1 \\
OneFlow      & 0.56 & 0.56 & 9.4  \\
MindSpore    & 3.90 & 3.90 & 11.6 \\
Chainer      & 0.32 & 0.32 & 8.5  \\
NumPy        & 1.64 & 1.64 & 9.2  \\
SciPy        & 4.84 & 4.84 & 12.4 \\
scikit-learn & 0.71 & 0.71 & 10.7 \\
OpenCV       & 1.08 & 1.08 & 11.8 \\
\midrule
\textbf{Total / Avg.} & 23.17 & 23.17 & 10.8 \\
\bottomrule
\end{tabular*}
\end{table}

\section{Discussion}
\label{sec:discussion}

\textbf{Supporting a New Library}.
Adding a library to \tool\ requires only a thin library-specific layer for documentation collection and input materialization; across the eleven libraries added beyond OpenCV, this layer requires a median of 112 non-comment lines of code, while the extraction prompt, \textit{ParamSpec} schema, relationship resolver, generation strategies, and test oracles remain unchanged. In contrast, FreeFuzz requires a usage corpus for each target and DocTer relies on library-specific documentation-parsing rules. The documentation-evolution experiment further shows that, once a library is supported, newly documented APIs can be tested without modifying the pipeline.

\textbf{Choice of Oracles}.
The crash and NaN oracles need no library-specific modeling and exploit the extracted constraints: a crash or a non-finite output inconsistent with documented behavior on a constraint-satisfying input signals a potential implementation--documentation inconsistency, and RQ1 shows the two oracles are complementary across library families. Their limitation is silent semantic errors---finite but incorrect results {where a finite result is expected}---which require semantic oracles such as cross-library differential testing~\cite{pham2019cradle,wang2022eagle}; the \textit{ParamSpec} representation already enables feeding equivalent inputs to corresponding APIs across libraries, providing a foundation for this next step.

\textbf{Choice of LLM}.
We replace the conference version's GPT-4 with a locally served \texttt{qwen2.5-coder:32b}~\cite{hui2024qwen} under greedy decoding (temperature $0$). We view this primarily as a reproducibility and deployability improvement: extraction becomes deterministic and independent of a proprietary service. The comparison in Section~\ref{sec:rq4} further shows that the smaller 7B variant reduces relationship-extraction recall and downstream VGR.

\textbf{Threats to Validity}.
\textit{Internal validity:} implementation bugs could distort results; we manually minimized and re-executed every reported issue, extraction is deterministic, hallucinations are caught by signature validation before generation (Section~\ref{sec:rq3}){, the effect of semantic extraction errors is measured in RQ4}, and all extracted \textit{ParamSpec}s are released for inspection.
Our oracles treat documentation as the specification, so documentation errors surface as false positives, which we exclude during triage.
{\textit{Construct validity:} bug detection is compared on a known-bug benchmark under the same crash and NaN oracles (RQ2c), which is restricted to bugs observable as crashes or non-finite outputs and cannot credit a baseline for bugs that only its own oracle, such as differential testing, would reveal. Excluding the baseline resources that are unusable in the evaluated versions may understate what the baselines support on their original versions.}
\textit{External validity:} findings are tied to twelve libraries, specific versions, and one extraction model, whose pretraining data may cover the tested libraries (the Sig-only configuration of RQ4 measures the dependence on the supplied documentation without ruling out such exposure); the served model's 4,096-token context bounds the documentation visible per API {(RQ4 reports its effect)}; and OpenCV was re-tested under the unified protocol, so its numbers are not comparable to the conference paper.

\textbf{Limitations}.
\tool\ executes three normalized relationship patterns rather than all possible cross-parameter constraints; in our manually labeled sample, 9.8\% of documented relationships fall outside this representation. Its crash and NaN oracles also cannot detect finite but semantically incorrect outputs {where a finite output is expected}. {Finally, \tool\ tests one API per invocation at the level of its arguments: APIs that lack usable documentation, need external resources, or need state set up through other APIs are outside its scope (Section~\ref{sec:evaluation}), and it does not generate API sequences.}

\section{Related Work}
\label{sec:related}

Fuzzing~\cite{miller1990empirical} executes the target system with random or invalid inputs to uncover anomalies and has been widely applied to operating systems~\cite{chen2022sfuzz}, network protocols~\cite{pham2020aflnet}, web applications~\cite{atlidakis2019restler}, and APIs~\cite{deng2022fuzzing}.

\textbf{Fuzzing Python Libraries}.
FreeFuzz~\cite{wei2022free} mines and mutates recorded API invocations, while DocTer~\cite{xie2022docter} extracts DL-specific structure, dtype, shape, and valid-value constraints from documentation through learned rules. Other work exercises libraries through generated neural networks~\cite{gu2022muffin,pham2019cradle,wang2020lemon,guo2020audee}, builds semantic oracles for numerical kernels~\cite{wang2022eagle,zhang2021predoo,yang2023fuzzing}, transfers inputs between relationally equivalent APIs~\cite{deng2022fuzzing}, or extracts constraints from operator source code~\cite{shi2023acetest}, which is exact for the checks the code performs but requires library-specific analysis of operator implementations and misses constraints the code does not check, such as documented numerical domains. FUTURE~\cite{li2025future} uses historical bugs of existing libraries and fine-tuned LLMs to test newly introduced libraries such as MindSpore and OneFlow. \tool\ differs primarily in scope and representation: it converts heterogeneous documentation and runtime signatures into a common \textit{ParamSpec} and applies the same normalized relationship-resolution and generation pipeline across twelve libraries.

{\textbf{Learning Library Constraints}. HOPPER~\cite{chen2023hopper} turns C library fuzzing into interpreter fuzzing and learns intra-API constraints, such as non-null pointers and minimum array lengths, from crash feedback and inter-API constraints, i.e., which calls must precede others, from coverage feedback. \tool\ instead obtains the relationships among the parameters of a single Python API from documentation before any execution and resolves them while generating each input; it does not synthesize API sequences.}

\textbf{LLM-Based Test Generation}.
TitanFuzz~\cite{deng2023large}, FuzzGPT~\cite{deng2024large}, and Fuzz4All~\cite{xia2024fuzz4all} generate and mutate complete test programs with LLMs, CHATAFL~\cite{meng2024large} applies LLMs to protocol fuzzing, LISP~\cite{li2024llm} to input-space partitioning, and further work to unit-test generation~\cite{schafer2024empirical,lemieux2023codamosa} and software engineering broadly~\cite{hou2024large}. These generation-based approaches rely on the LLM's prior exposure to idiomatic usage of the target library, so generated programs frequently violate input constraints on less-represented libraries. \tool\ makes the constraints explicit by extracting them from the supplied documentation and invokes a locally served open-sourced model~\cite{hui2024qwen} only once per API, incurring no per-test inference cost and no dependence on proprietary services.

{\textbf{LLM-Based Constraint Extraction}. The conference version of \tool~\cite{duan2025harnessing} already used an LLM to extract parameter constraints and inter-parameter dependencies from API documentation. FD-FACTORY~\cite{zheng2026fdfactory}, published afterwards, uses an LLM-based agent to turn documentation and usage code into a constraint report, including inter-parameter dependencies, mutual exclusions, and conditional requirements; further LLM steps then synthesize, validate, and repair a fuzz driver for each API, which is fuzzed with Atheris without further LLM calls. FD-FACTORY thus realizes the constraints in an LLM-written driver dedicated to each API, whereas \tool\ keeps them in a fixed, inspectable schema, validated against the runtime signature, that one generic generator executes across libraries: a new API needs a specification rather than driver code, and a new library only the adapter of Section~\ref{sec:rq4}. Conversely, FD-FACTORY's constraint report can express mutual exclusions and conditional requirements, which \tool's three patterns do not cover (see Limitations in Section~\ref{sec:discussion}). Other recent work obtains constraints from execution or code rather than documentation: DeepConstr~\cite{go2024deepconstr} refines operator constraints by checking whether their complement still yields valid tests, Centaur~\cite{naziri2026centaur} lets an LLM enumerate candidate constraint rules, validates them on seed inputs, and solves them with an SMT solver, and DLLens~\cite{li2024dllens} extracts path constraints from API implementations with LLM-aided static analysis. These approaches are evaluated on PyTorch and TensorFlow, whereas \tool\ obtains constraints from documentation with one pipeline across twelve libraries; our comparison includes ACETest as the representative of constraint extraction from code.}

\textbf{Document-Guided Testing}.
Earlier document-guided techniques detect specification--implementation inconsistencies~\cite{lv2020rtfm,zhou2018automatic}, derive assertions and oracles~\cite{liu2014automatic,motwani2019automatically}, or extract executable constraints through manually designed or learned rules~\cite{blasi2018translating,xie2022docter}. In particular, DocTer also {resolves the generation order of dependent parameters} through a dependency graph{, but extracts dependencies with rules learned for specific documentation formats}. \tool\ normalizes recurring relationships into three explicit executable patterns, shape following, rank-bounded axis, and type following, within a common representation used across heterogeneous Python libraries. Across our twelve-library evaluation, 40.1\% of tested APIs contain at least one such relationship, and disabling their resolution reduces VGR on these APIs from above 95\% to 32.1\%--52.3\%.

\section{Conclusion}
\label{sec:conclusion}

This paper introduced \tool, a document-guided fuzzing approach for Python library APIs. \tool\ uses a locally served open-sourced LLM to parse heterogeneous API documentation into standardized parameter specifications, validates them against runtime signatures, and generates test inputs that satisfy both individual parameter constraints and inter-parameter dependencies. Applied to 7,718 APIs across twelve Python libraries, \tool\ detected 74 issues, of which 47 have been confirmed by developers as real bugs and 38 have already been fixed. On PyTorch and TensorFlow, we compared it with API-level and LLM-based fuzzers and with the same LLM prompted to generate test inputs directly, separating API applicability, per-API effectiveness, and bug detection: it supports more APIs and covers more lines than the API-level fuzzers, is comparable with TitanFuzz on the commonly supported APIs, on which the same LLM prompted directly covers 23\% and 25\% fewer lines, and detects 12 of 27 known bugs, as many as TitanFuzz and more than the other baselines. The same core pipeline is applied across all twelve libraries.

\balance
\bibliographystyle{IEEEtran}
\bibliography{main}

\end{document}